\documentclass[traditabstract]{aa}
\usepackage{savesym}
\usepackage{amsmath}
\savesymbol{iint}
\usepackage{txfonts}
\restoresymbol{TXF}{iint}
\usepackage{graphicx}
\usepackage{amsfonts}
\usepackage{amssymb}
\usepackage{epsfig}
\usepackage{color}
\usepackage{epsf}
\usepackage{longtable,lscape}
\usepackage{rotating}
\usepackage{thumbpdf}
\usepackage{natbib}
\usepackage{color}

\begin{document}

\title{
Particle transport in intense small-scale magnetic turbulence with a mean field
}
\author{
        I. Plotnikov\inst{1} \and G.
        Pelletier\inst{1} \and  M. Lemoine\inst{2} }
\institute{UJF-Grenoble 1/CNRS-INSU, Institut de Planétologie et d'Astrophysique de Grenoble (IPAG) UMR 5274, 38041 Grenoble, France
\and Institut d'Astrophysique de Paris, CNRS -- UPMC, 98 bis boulevard
Arago, 75014 Paris, France.}

\date{}    

\abstract{Various astrophysical studies have motivated the
  investigation of the transport of high energy particles in magnetic
  turbulence, either in the source or en route to the observation
  sites. For strong turbulence and large rigidity, the pitch-angle
  scattering rate is governed by a simple law involving a mean free
  path that increases proportionally to the square of the particle
  energy. In this paper, we show that perpendicular diffusion deviates
  from this behavior in the presence of a mean field. We propose an
  exact theoretical derivation of the diffusion coefficients and show
  that a mean field significantly changes  the transverse diffusion
  even in the presence of a stronger turbulent field. In particular,
  the transverse diffusion coefficient is shown to reach a finite
  value at large rigidity instead of increasing proportionally to
  the square of the particle energy. Our theoretical derivation is
  corroborated by a dedicated Monte Carlo simulation.  We briefly
  discuss several possible applications in astrophysics.}

\keywords{Magnetic fields -- Turbulence -- Diffusion}

\titlerunning{Particle transport in intense small-scale magnetic turbulence}

\maketitle

\section{Introduction}
The scattering and the spatial diffusion of high energy particles off
magnetic turbulence play a crucial role in many fields of
astrophysics. For instance they are key ingredients of Fermi acceleration
processes because they directly control the efficiency and the rate of
particle acceleration. They determine the properties of the confinement of
astrophysical objects from jets to galaxies and clusters of galaxies,
and governs the transport of the particles through interplanetary,
interstellar, or intergalactic space. Diffusion has long
been described by a quasi-linear theory approach (Jokipii 1966,
1973), which allows us to calculate the diffusion coefficients when the
turbulent field is significantly weaker than the background
field. However, in many circumstances the level of turbulence turns
out to be large so that this standard picture requires
extension. Several studies have examined the transport properties in
strong turbulence by means of numerical simulations, e.g.  Giacalone \&
Jokipii (1999), Casse et al. (2002), Candia \& Roulet (2004), and Fatuzzo et al. (2010).
  Most of these investigations have focused on the
situation in which a large-scale turbulence cascades toward small
dissipative scales -- as in the Kolmogorov scheme -- and in which
particles interact with gyroresonant modes of the turbulence
spectrum. From the point of view of the particle, the turbulence
therefore occurs on large scales, as the coherence length of the
magnetic field corresponds roughly to the maximal scale of the turbulent
spectrum.

However, in a variety of physical situations, the Larmor radius of the
particle can exceed the coherence scale of the turbulence. The
transport of particles downstream of a relativistic shock wave
provides a clear example of this situation. The mean field is there
mostly transverse to the flow because of Lorentz tranform effects and
shock compression, and the turbulence that is excited in the shock
precursor is generated on microscopic plasma skin depth scales. In
this case, perpendicular diffusion at high (possibly very high)
rigidity controls the transport of the particles back and forth from
the shock. More generally, the high rigidity regime likely plays an
important role in the deconfinement process of particles of high
energy, when their Larmor radius exceeds the size of the
astrophysical system. However, this high rigidity regime has received
little attention so far, except for the pioneering study of Shalchi \&
Dosch (2009). The pitch angle scattering rate is known to increase in
proportion to the square of the particle energy in this limit, but the
behavior of the transverse diffusion coefficient, which is crucial in
the above contexts deserves a careful analysis. This analysis is the
objective of the present paper. It will be found in particular that
even a weak mean field, as measured relatively to the turbulent
component, can affect the scaling of the perpendicular diffusion
coefficient.

The present paper describes both a theoretical and a numerical study of
diffusion at high rigidity. The theoretical aspects are discussed in
Section~\ref{sec:theor}, while the numerical simulations are presented
in Section~\ref{sec:num}. Finally in Section~\ref{sec:appl} we summarize
our results and discuss some applications.

\section{Transport of high rigidity particles with a mean field}\label{sec:theor}

\subsection{Notations and summary of previous results}
The transport of particles in magnetostatic turbulence is
characterized by the reduced rigidity $\rho$, the level of turbulence
$\eta$, and the power spectrum of magnetic fluctuations in three dimensions (hereafter 3D) $S_{\rm
  3d}(\mathbf{k})$. These quantities are defined as
\begin{equation}
\rho \equiv \frac{\bar r_{\rm L}}{\ell_{\rm c}} =
\frac{\epsilon}{e\bar B \ell_{\rm c}} \ ,
\end{equation}
where $\bar r_{\rm L}$ denotes the Larmor radius of the particle in
the total (mean $\mathbf{B_0}$ and turbulent $\boldsymbol{\delta B}$)
field $\bar B$ where $\boldsymbol{\bar B}^2\,\equiv\,\boldsymbol{B_0}^2 +
\boldsymbol{\delta B}^2$, $\epsilon$ the energy of the particle, and
$\ell_{\rm c}$ the coherence length of the fluctuations. 

The turbulence level $\eta$ is defined as
\begin{equation}
  \eta\,\equiv\,\frac{\langle \boldsymbol{\delta
      B}^2\rangle}{\langle \boldsymbol{\delta
      B}^2\rangle + \boldsymbol{B_0}^2}\ ,
\end{equation}
where $\eta\rightarrow 0$ corresponds to weak turbulence and
$\eta\rightarrow 1$ corresponds to pure turbulence with no mean field.

The correlation function $C(\mathbf{r})$ of the random field
\begin{equation}
  C(\mathbf{r})\,\equiv\, \frac{\langle \boldsymbol{\delta
      B}(\boldsymbol{x}+\boldsymbol{r})\boldsymbol{\delta
      B}(\boldsymbol{x})\rangle}{\langle \boldsymbol{\delta
      B}^2\rangle}\ ,
\end{equation}
can be written in terms of the one-dimensional power spectrum $S(k)
\propto k^2 S_{\rm 3d}(\mathbf{k})$
\begin{equation}
C(r)\,=\,\frac{\int\mathrm{d}k\,S(k)\,\mathrm{sin}(kr)/(kr)}
{\int\mathrm{d}k\,S(k)}\ .\label{eq:corr-func}
\end{equation}
 Casse et al. (2002) defined the coherence length as the
scale at which $C(r)$ is maximum; if the power spectrum takes the form
of a broad-band truncated power-law $S(k) \propto (k/k_{\rm
  min})^{-\beta}$ for $k_{\rm min}\leq k\leq k_{\rm max}$ and zero
otherwise, one finds for the coherence length $\ell_{\rm c}\simeq 0.77
k_{\rm min}^{-1}$. Alternatively, one can define the coherence length
as we do here, to be
\begin{equation}
\ell_{\rm c}\,\equiv\, \int_{0}^{+\infty}{\rm d}r\,C(r)\ , \label{eq:corr-length}
\end{equation}
 where one then derives in a straightforward way
\begin{equation}
\ell_{\rm c}\,=\,\frac{\pi}{2}\frac{1}{\eta}\int_0^{+\infty} {\rm
  d}k\,k^{-1}\,S(k)\ ,
\end{equation}
and the presence of $1/\eta$ results from our choice of normalization
for the power spectrum 
\begin{equation}
\int_0^{+\infty} {\rm d}k\,S(k)\,\equiv\,\eta\ ,
\end{equation}
where in practice the spectrum is bounded between $k_{\rm min}$ and $k_{\rm max}$.
Both definitions for $\ell_{\rm c}$ coincide to within a factor close to
unity.  As a function of the spectrum index $\beta$, the coherence length is close to either 
$k_{\rm min}^{-1}$ on larges scale for $\beta>1$, or to $k_{\rm max}^{-1}$ on small 
scales for $\beta<1$.

The scattering frequency $\nu_{\rm s}$ is defined as the
reciprocal of the decorrelation time of the pitch angle of the
particle, the latter being defined relative to the direction of the
mean field. As discussed in Casse et al. (2002), the scattering
frequency can be written
\begin{equation}
  \nu_{\rm s} \,\approx\,  \frac{\pi}{3} \frac{c}{\bar r_{\rm
      L}^2}\frac{\int_{k\bar r_{\rm L} >1}\, k^{-1}S(k)\, {\rm
        d}k}{\int S(k)\,{\rm d}k}  \ ,
\end{equation}
an expression that extends to the strong turbulence regime the results
of the quasi-linear theory. This leads to the scalings
\begin{eqnarray}
\nu_s & \simeq & \frac{2}{3}\eta\frac{c}{\ell_{\rm c}} \rho^{\beta-2}
 \quad (\rho \ll 1) \nonumber\\
\nu_s & \simeq & \frac{2}{3}\eta \frac{c}{\ell_{\rm c} \rho^2} \quad
(\rho \gg 1)\ .
\end{eqnarray}
The Bohm scaling holds only in the very special case where $\beta=1$.
In addition to these quantities, the notion of correlation time also plays
an important role because it measures the time beyond which a particle
experiences a force that is decorrelated from the initial one, along
the particle trajectory. It is then defined as
\begin{equation}
\label{eq:corr-time}
\tau_{\rm c} \equiv \int_0^{+\infty} C(\vert \Delta x(\tau) \vert) d\tau \ ,
\end{equation}
where $\Delta x(\tau)$ represents the displacement after a time $\tau$
in the turbulence. In quasi-linear theory, only the unperturbed
trajectory is inserted into this definition, although one can extend that
definition with a diffusive trajectory as we later indicate.

If a relativistic particle travels over a coherence length of the
turbulent field without having displayed any wiggle, corresponding to
the regime $\rho\gg1$, then $\tau_{\rm c} \sim \ell_{\rm c}/c$. This
correlation time is much shorter than the scattering time $\nu_{\rm
  s}^{-1}\sim \eta^{-1}\rho^2 \ell_{\rm c}/c$ in this regime. The
correlation time $\tau_{\rm c}$ can be recovered from
Eq.~(\ref{eq:corr-time}) by using the ballistic approximation $\Delta
x(\tau)\simeq c\tau$, which is appropriate in this regime $\rho\gg1$,
in which case Eq.~(\ref{eq:corr-func}) leads to $\tau_{\rm
  c}=(\pi/2)(\beta-1)\beta^{-1}k_{\rm min}^{-1}/c\sim \ell_{\rm c}/c$.
 We note that in the special case where the power-law index of turbulence $\beta=1$ (Bohm regime) Eqs.~(\ref{eq:corr-func}) 
 and (\ref{eq:corr-length}) lead to $ \ell_{\rm c} = ({\lambda_{\rm {min}} / 4) \log ({\lambda_{\rm {max}} / \lambda_{\rm{min}}}) }$, where $\lambda_{\rm min}$ and $\lambda_{\rm max}$ are the shortest and the longest wavelengths of turbulence.

If a particle experiences a chaotic motion on a length-scale smaller than
$\ell_{\rm c}$, corresponding to the regime $\rho\ll1$, then the
estimate is more complicated to obtain but one finds that $\tau_{\rm c}
\sim \rho^{\beta}\ell_{\rm c}/c$ as follows. Since the
correlation time remains shorter than the scattering time, Casse et
al. (2002) proposed a heuristic estimate in which decorrelation
arises out of the small-scale modes with wavenumber $k > k_{\rm
  min}\rho^{-1}$, which give rise to gyroresonant interactions with
the particle of rigidity $\rho$. The modes with wavelengths longer
than the Larmor radius (i.e. $k<k_{\rm min}\rho^{-1}$) construct the
field line to which the particle is attached, hence do not cause
 decorrelation on timescales shorter than the scattering time.  The
above correlation time is indeed shorter than the scattering time and
 increases with $\rho$. The heuristic estimate for $\rho <1$ is consistent
  with quasi-linear theory when $\eta \ll 1$ and with numerical
results in the strong turbulence regime (Casse et al. 2002) can then be
written as
\begin{equation}
\tau_{\rm c} \simeq  \frac{1}{\eta c}\int_{k>k_{\rm min}\rho^{-1}} {\rm d}k\,
  k^{-1} S(k) \ ,
\end{equation}
which bears some resemblance to the case discussed before for
$\rho\gg1$, except that $\rho$ explicitly enters the sinc function,
since one must now follow the orbit of the particle around the field line,
and the integral is limited to $k>k_{\rm min}\rho^{-1}$ for the
reasons given above. The calculation then implies that $\tau_{\rm c} \sim 
\rho^{\beta}\ell_{\rm c}/ c$ as announced.  The particle trajectory
undergoes decoherence before traveling $\ell_{\rm c}$ because of the
large number of wiggles in the random field.

Thus, except for $\eta \sim 1$ and $\rho \sim 1$ for which the
correlation time becomes comparable to the scattering time, a
Markovian theory of the scattering process is appropriate, even if the
turbulence is strong, stronger even than the mean field. This is an
essential key for the present discussion.

Independently of the rigidity, the parallel diffusion coefficient is
always given by $D_{\parallel} = c^2/(3\nu_{\rm s})$, even
in the strong regime of turbulence. As for the transverse diffusion
coefficient, in the strong regime at low rigidities, it does not
follow a law similar to the quasi-linear result but is proportional to
$D_{\parallel}$ (Casse et al. 2002) because of the magnetic field
line wandering that transmits parallel diffusion in the transverse
direction.  Casse et al. (2002) found in particular that $D_{\perp}
= \eta^{2.3} D_{\parallel}$ at small rigidities, which rules out the
conjecture of Bohm's diffusion. In the next section, we discuss the
transverse diffusion in the large rigidity regime.\\

\subsection{Transverse diffusion at large rigidity}

As mentioned previously, in the large rigidity regime $\rho\gg1$, the
correlation time is (much) shorter than the scattering time, hence we
 expect to derive the parallel and transverse diffusion
coefficients using a Markovian description of the trajectory. In
particular when $\rho\gg1$, the velocity changes by $1/\rho$ only over
a correlation time. This implies that significant changes in the
velocity occur on timescales that are much longer than the correlation
time. Therefore we can assimilate the effect of small-scale
fluctuations to a fully decorrelated white noise on the relevant
timescales.

To calculate the particle transport in a random field, one
has to use the solution of the differential equation that
governs the evolution of the particle velocity $\mathbf v$
\begin{equation}
\frac{\rm d}{{\rm d}t} {\mathbf v} = \left[{\boldsymbol {\hat \Omega_0}} +
  {\boldsymbol {\delta \hat \Omega(t)}}\right]
 \cdot {\mathbf v}\ .
\end{equation}
The quantities ${\boldsymbol {\hat \Omega_0}}$ and ${\boldsymbol
  {\delta \hat \Omega(t)}}$ are rotation operators developed as linear
combinations of the generators of the Lie algebra of the rotation
group, $\boldsymbol{\hat L_1}$, $\boldsymbol{\hat L_2}$, $\boldsymbol{\hat L_3}$
\begin{equation}
  \hat L_1\, := \begin{pmatrix}
    0 & 0 & 0 \\
    0 & 0  & -1 \\
    0 &  1  & 0 \\
\end{pmatrix} \ , \hat L_2\, := \begin{pmatrix} 0 & 0 & 1 \cr 0 & 0 &
  0\cr -1 & 0 & 0 \cr
\end{pmatrix} \ , \hat L_3\, := \begin{pmatrix} 0 & -1 & 0 \cr 1 & 0 &
  0 \cr 0 & 0 & 0 \cr
\end{pmatrix} \ .
\end{equation}
In detail, $\boldsymbol{\hat\Omega_0}=\Omega_0 B_0^i \boldsymbol{\hat
  L_i}/B_0$, where $B_0^i$ denotes the $i$-th component of
$\mathbf{B_0}$ and $\Omega_0\equiv c/r_{\rm L,0}$ the Larmor pulsation
defined with respect to the mean field.  With this notation,
${\boldsymbol {\hat \Omega_0}} \cdot {\mathbf v} = \Omega_0
\mathbf{v}\times\mathbf{B_0}/B_0$. The operator ${\boldsymbol {\delta
    \hat \Omega(t)}}$ is decomposed in a similar way as the generators
of the rotation group, and $\delta \Omega \equiv c/r_{\rm L}$, where
$r_{\rm L}$ is now measured relatively to $\delta B$.

To solve the equation of motion, one uses an auxiliary variable
$\mathbf{u}$ that is defined as
\begin{equation}
\mathbf{v(t)}\,\equiv\, {\boldsymbol{\hat R_0(t)}}\cdot\mathbf{u(t)},
\end{equation}
where
\begin{equation}
{\boldsymbol{\hat R_0(t)}}\,\equiv\,\exp\left(t{\boldsymbol {\hat
      \Omega_0}}\right)\ .
\end{equation}
We then define
\begin{equation} 
  \boldsymbol {\hat {\tilde \Omega}(t)} \equiv
    \boldsymbol {\hat R_0(t)^{-1}}\cdot \boldsymbol {\delta \hat
        \Omega(t)}\cdot \boldsymbol{\hat R_0(t)} \ ,
\end{equation}
one finds that $\mathbf{u(t)}$ obeys
\begin{equation}
\frac{\rm d}{{\rm d}t}{\mathbf u} = \boldsymbol{\hat{\tilde \Omega}(t)} \cdot
{\mathbf u}\ .
\end{equation}
This equation is solved as 
\begin{equation}
\mathbf{u}(t)\,=\,{\cal T} \exp\left[\int_0^t \boldsymbol{\hat{\tilde
    \Omega}(t')}\, {\rm d}t'\right] \cdot \mathbf{u}(0) \ .
\end{equation}
Because the operator in the exponent is time dependent, to preserve the exponential character of the solution, a time-ordering operator $\cal{T}$ has to be introduced, as we now explain.

We note that $\mathbf{u}(0)=\mathbf{v}(0)$, thus the solution for
$\mathbf{v}$ is given by
\begin{equation}
\mathbf{v}(t)\,=\,\boldsymbol{\hat R_0(t)}\cdot {\cal T} \exp\left[\int_0^t \boldsymbol{\hat{\tilde
    \Omega}(t')}\, {\rm d}t'\right] \cdot \mathbf{v}(0) \ .
\label{eq:sol-v}
\end{equation}
The regular part of the field generates the regular rotation matrix
$\boldsymbol{\hat R_0(t)}$, while the exponential accounts for the
effect of the turbulent part. The time-ordering operator ${\cal T}$
maintains the chronological order of the products in the non-commuting
$\hat{\tilde \Omega}(t_k)$ in the expansion of the exponential
operator, i.e.
\begin{eqnarray} 
{\cal T} \boldsymbol{\hat \Omega(t_1)} \cdot\boldsymbol{ \hat \Omega(t_2)} &=&
  \boldsymbol{\hat \Omega(t_1)} \cdot \boldsymbol{\hat \Omega(t_2)} \, \, {\rm if} \, \,
  t_1>t_2\ , \nonumber\\
&=&  \boldsymbol{\hat \Omega(t_2)} \cdot\boldsymbol{ \hat \Omega(t_1)} \, \,
  {\rm if} \, \, t_2>t_1 \ ,
\end{eqnarray}
and so on for higher order products. Alternatively, the time-ordered
expansion can be written as a Dyson series
\begin{eqnarray}
{\cal T} \exp\left[\int_0^t \boldsymbol{\hat{\tilde
    \Omega}(t')}\, {\rm d}t'\right]&&\,\equiv\,
1 + \nonumber\\
&& \sum_{n=1}^{n=+\infty} \int_{0}^{t}{\rm
  d}t_1\ldots\int_{0}^{t_{n-1}}{\rm d}t_n \,\boldsymbol{\hat{\tilde
    \Omega}(t_1)}\ldots\boldsymbol{\hat{\tilde
    \Omega}(t_n)}\ .\nonumber\\
&&\label{eq:Dyson-series}
\end{eqnarray}

We now use the following theorem that holds for a Gaussian stationary
random process in the {\it white noise limit}. As discussed in detail
in the Appendix, this is a direct generalization to any Lie algebra of
a well-known result for a scalar random process, with no other
restriction than the white noise assumption
\begin{eqnarray}
  &&\left\langle{\cal T} \exp\left[ \int_0^t \boldsymbol{\hat {\tilde \Omega}
        (t')}\, {\rm d}t' \right]\right\rangle =\nonumber \\ 
  && \quad {\cal T}\exp\left[\frac{1}{2}
    \int_0^t {\rm d}t_1  \int_0^t {\rm d}t_2\, \left\langle
      \boldsymbol{\hat {\tilde \Omega}(t_1)}
 \cdot \boldsymbol{\hat{\tilde \Omega}(t_2)}\right\rangle\right] \ .
\label{eq:av-Tord}
\end{eqnarray}
Various properties of the turbulent field can be considered i.e. that is either isotropic
with no helicity, isotropic with helicity, or anisotropic with rotation
invariance in the transverse direction. All these cases can be easily
treated, although we focus on two relevant cases: (A)
3D isotropic turbulence and (B) two-dimensional (hereafter 2D)
isotropic turbulence in the plane transverse to $\mathbf{B_0}$, with
$\boldsymbol{\delta B}\cdot\mathbf{B_0} =0$. 

We define the projection operators $\boldsymbol{\hat \pi_{\perp}}$ on
  the plane transverse to $\mathbf{B_0}$ and the projection operator
  $\boldsymbol{\hat \pi_{\parallel}}$ along $\mathbf{B_0}$. We now
    define the correlation function of the random rotation matrices
    $\boldsymbol{\hat{\delta \Omega}}$
\begin{equation}
\langle\boldsymbol{\hat{\delta \Omega}(t_1)}\boldsymbol{\hat{\delta
    \Omega}(t_1)}\rangle\,=\,
\langle \delta\Omega^i(t_1)\delta\Omega^j(t_2)\rangle\boldsymbol{\hat L_i}\boldsymbol{\hat L_j}\
,
\end{equation}
where
\begin{equation}
\langle \delta\Omega^i(t_1)\delta\Omega^j(t_2)\rangle\,=\,2\tau_{\rm
  c}\delta(t_1-t_2)\,\left[\frac{1}{2}\langle\delta\Omega_\perp^2\rangle\boldsymbol{\hat
    \pi_\perp}^{ij} +
    \langle\delta\Omega_\parallel^2\rangle\boldsymbol{\hat\pi_\parallel}^{ij}\right]\ .
\end{equation}
The scalars $\langle\delta\Omega_\parallel^2\rangle$ and
$\langle\delta\Omega_\perp^2\rangle$ characterize the relative
strengths of the turbulence in the parallel (to $\mathbf{B_0}$) and
perpendicular directions. In particular, for 3D isotropic turbulence,
$\langle\delta\Omega_\perp^2\rangle=2\langle\delta\Omega_\parallel^2\rangle$,
in which case the above correlator becomes proportional to the
identity.  Then, using the properties of $\boldsymbol{\hat\pi_\perp}$,
$\boldsymbol{\hat\pi_\parallel}$ and the $\boldsymbol{\hat L_i}$, one finds
\begin{eqnarray}
  \langle\boldsymbol{\hat{\delta \Omega}(t_1)}\boldsymbol{\hat{\delta
      \Omega}(t_1)}\rangle &\,=\,& -2\tau_{\rm c}\delta(t_1-t_2)
\nonumber\\
&&\,\,\times\,\left[
    \langle\delta\Omega^2\rangle\boldsymbol{\hat{1}} -
    \langle\delta\Omega_\parallel\rangle^2\boldsymbol{\hat\pi_\parallel}
    -\frac{1}{2}\langle\delta\Omega_\perp^2\rangle\boldsymbol{\hat\pi_\perp}\right]\ ,
\end{eqnarray}
where $\langle\delta\Omega^2\rangle\equiv
\langle\delta\Omega_\parallel\rangle^2+
\langle\delta\Omega_\perp^2\rangle$.  We note that the above correlation
function holds for $\boldsymbol{\hat{\delta \Omega}}$, which should not
  be confused with $\boldsymbol{\hat{\tilde \Omega}}$, the latter being
    the quantity of relevance for calculating the transport
    properties, as expressed in Eq.~(\ref{eq:av-Tord}). However,
\begin{equation}
\langle\boldsymbol{\hat{\tilde \Omega}(t_1)}\boldsymbol{\hat{\tilde
    \Omega}(t_2)}\rangle
= e^{-t_1\boldsymbol{\hat \Omega_0}}\langle\boldsymbol{\hat{\delta
    \Omega}(t_1)}e^{-(t_2-t_1)\boldsymbol{\hat \Omega_0}}
\boldsymbol{\hat{\delta
    \Omega}(t_2)}\rangle e^{t_2\boldsymbol{\hat \Omega_0}}\ ,
\end{equation}
and, because $\left[\boldsymbol{\hat\pi_\parallel},e^{t\boldsymbol{\hat
      \Omega_0}}\right]=\left[\boldsymbol{\hat\pi_\perp},e^{t\boldsymbol{\hat
      \Omega_0}}\right]=0$, the correlation function for
$\boldsymbol{\hat{\tilde \Omega}}$ is the same as that for
$\boldsymbol{\hat{\delta \Omega}}$.

Using Eq.~(\ref{eq:av-Tord}), one then finds the solution for
$\mathbf{v}$:
\begin{equation}
\langle\mathbf{v}(t)\rangle\,=\,\boldsymbol{\hat R_0(t)}\cdot \exp\left[\frac{1}{2}\int_{0}^t{\rm
    d}t_1 \int_{0}^t{\rm d }t_2\,\langle\boldsymbol{\hat{\tilde \Omega}(t_1)}\boldsymbol{\hat{\tilde
    \Omega}(t_2)}\rangle\right]\mathbf{v}(0)\ ,
\end{equation}
where the average is taken over the possible realizations of the
turbulent field. This leads to
\begin{equation}
\langle\mathbf{v}(t)\rangle\,=\,\boldsymbol{\hat
  R_0(t)}\cdot\exp\left[-t\tau_{\rm
    c}\left(\langle\delta\Omega^2\rangle 
- \langle\delta\Omega_\parallel^2\rangle\boldsymbol{\hat\pi_\parallel}-
\frac{1}{2}\langle\delta\Omega_\perp^2\rangle
\boldsymbol{\hat\pi_\perp}\right)\right]\mathbf{v}(0)\ .
\end{equation}
Using the properties of $\boldsymbol{\hat\pi_\parallel}$ and $\boldsymbol{\hat\pi_\perp}$, this
can be rewritten as
\begin{eqnarray}
  \langle\mathbf{v}(t)\rangle&\,=\,&
  \boldsymbol{\hat  R_0(t)}\cdot\Biggl\{\exp\left[-t\tau_{\rm
      c}\left(\langle\delta\Omega^2\rangle-\langle\delta\Omega_\parallel^2\rangle\right)\right]\boldsymbol{\hat\pi_\parallel}\nonumber\\
  &&    + \exp\left[-t\tau_{\rm
      c}\left(\langle\delta\Omega^2\rangle-\frac{1}{2}\langle\delta\Omega_\perp^2\rangle\right)\right]\boldsymbol{\hat\pi_\perp}\Biggr\}\mathbf{v}(0)\
  .
\end{eqnarray}
Therefore, one derives the general results
\begin{equation}
\langle v_\parallel(0) v_\parallel(t)\rangle \,=\, \exp\left[-t\tau_{\rm
    c}\left(\langle\delta\Omega^2\rangle-\langle\delta\Omega_\parallel^2\rangle\right)\right]
\langle   v_\parallel(0)^2\rangle\label{eq:parv-corr}
  .
\end{equation}
In the transverse direction,
\begin{eqnarray}
  \langle\boldsymbol{v_\perp}(0)\cdot\boldsymbol{v_\perp}(t)\rangle&=&
  \exp\left[-t\tau_{\rm
      c}\left(\langle\delta\Omega^2\rangle-\frac{1}{2}\langle\delta\Omega_\perp^2\rangle\right)\right]\,\nonumber\\
&&\quad\quad\times\,  ^{\rm T}\boldsymbol{v_\perp}(0)\cdot\boldsymbol{\hat
    R_0}\cdot\boldsymbol{v_\perp}(0)\nonumber\\
  &=& \exp\left[-t\tau_{\rm
      c}\left(\langle\delta\Omega^2\rangle-\frac{1}{2}\langle\delta\Omega_\perp^2\rangle\right)\right]\nonumber\\
&&\quad\quad\times\,\cos\left(\Omega_0 t\right)
  \,\langle \boldsymbol{v_\perp}(0)^2\rangle\ .\label{eq:perpv-corr}
\end{eqnarray}
The last equality follows from developing the exponential
$\boldsymbol{\hat R_0}=\exp\left(t\boldsymbol{\hat \Omega_0}\right)$,
noting that $\boldsymbol{\hat \Omega_0}=\Omega_0 \boldsymbol{\hat
  L_3}$ for $\mathbf{B_0}$ oriented along $z$, $\boldsymbol{\hat
  L_3}^{2n}=(-1)^n\boldsymbol{\hat\pi_\perp}$, $\boldsymbol{\hat
  L_3}^{2n+1}=(-1)^n\boldsymbol{\hat L_3}$,
$^{\rm T}\boldsymbol{v_\perp}(0)\cdot\boldsymbol{\hat\pi_\perp}
\cdot\boldsymbol{v_\perp}(0)=\langle \boldsymbol{v_\perp}(0)^2\rangle$
, and $^{\rm T}\boldsymbol{v_\perp}(0)\cdot\boldsymbol{\hat L_3}
\cdot\boldsymbol{v_\perp}(0)=0$.

The parallel $D_\parallel$ and perpendicular $D_\perp$ diffusion coefficients are directly
obtained from the correlation functions of the velocity components
after averaging over the initial velocities
\begin{eqnarray}
  D_\parallel&\,=\,& \int_0^{+\infty}
  {\rm d}t\, \langle v_\parallel(0)v_\parallel(t)\rangle\ ,\nonumber\\
  D_\perp&\,=\,& \frac{1}{2}\int_0^{+\infty}
  {\rm d}t\, \langle
  \boldsymbol{v_\perp}(0)\cdot\boldsymbol{v_\perp}(t)\rangle\ .
\end{eqnarray}
Using Eqs.~(\ref{eq:parv-corr}) and (\ref{eq:perpv-corr}), this leads
to
\begin{eqnarray}
D_\parallel&\,=\,& \frac{1}{3}\frac{c^2}{\nu_{\rm s}}\ ,\nonumber\\
D_\perp&\,=\,& \frac{1}{3}c^2\frac{\nu_\perp}{\nu_\perp^2+\Omega_0^2}\
, \label{eq: coeff-dperp}
\end{eqnarray}
where
\begin{eqnarray}
\nu_{\rm s}&\,=\,& \tau_{\rm
    c}\left(\langle\delta\Omega^2\rangle-\langle\delta\Omega_\parallel^2\rangle\right)\
  ,
\nonumber\\
\nu_\perp&\,=\,& \tau_{\rm c}\left(\langle\delta\Omega^2\rangle-\frac{1}{2}
      \langle\delta\Omega_\perp^2\rangle\right)\ .
\end{eqnarray}
These expressions for $D_\perp$ are formally similar to the results of the
so-called classical diffusion theory, although they are obtained here under
different physical assumptions; in particular, a strong turbulence
situation is assumed. 

In case (A), for 3D isotropic turbulence,
$\langle\delta\Omega_\perp^2\rangle=
2\langle\delta\Omega_\parallel^2\rangle=\frac{2}{3}\langle\delta\Omega^2\rangle$,
so that 
\begin{equation}
  \nu_\perp=\nu_{\rm s}= \frac{2}{3}\langle\delta\Omega^2\rangle \tau_{\rm c} =
  \frac{2}{3}\frac{\eta}{\rho^2}\frac{c}{\ell_{\rm c}} \ .
\end{equation}
One may note that the expression for $\nu_{\rm s}$ matches that
derived from a random walk argument for pitch angle diffusion. We also note
that the above calculation for $\nu_{\rm s}$ may be applied to
the regime $\rho\ll1$, as long as the correlation time is shorter than
the scattering time. This is true in the case of $\nu_{\rm
  s}=(2/3)\eta\rho^{\beta-2}$, which is the standard quasi-linear
theory result. The result for the perpendicular coefficient
cannot, of course, be extended to the regime $\rho\ll 1$, as the above calculation
does not account for field line wandering.

In case (B), for 2D transverse isotropic turbulence,
$\langle\delta\Omega_\parallel^2\rangle=0$, 
$\langle\delta\Omega_\perp^2\rangle=\langle\delta\Omega^2\rangle$,
hence 
\begin{equation}
\nu_{\rm s}=2\nu_\perp=\langle\delta\Omega^2\rangle \tau_{\rm c}\ .
\end{equation}
This demonstrates that the transverse diffusion coefficients follows the
scalings, which we express here for case (A), i.e. isotropic turbulence
\begin{eqnarray}
D_\perp &\simeq& D_\parallel\,\simeq\,\frac{1}{2}c\ell_{\rm c}\rho^2/\eta\quad
(1\ll\rho \ll \bar B/B_0)\ ,\nonumber\\
D_\perp &\simeq& \frac{2}{9}c\ell_{\rm c} \frac{\bar B^2}{B_0^2}\quad
(\bar B/B_0\ll\rho)\ .\label{eq:dperp_th}
\end{eqnarray}
The transition between these two regimes takes place at $\rho \sim
\bar B/B_0\simeq \eta^{1/2}(1-\eta)^{-1/2}$, corresponding to
$\nu_{\rm s}\sim \Omega_0$. At larger rigidities, the perpendicular
diffusion coefficient remains constant, while the parallel diffusion
coefficient continues to increase as $\rho^2$.

This  result is supported by the numerical
simulation that we now discuss.

\section{Numerical simulation of the transport with a mean field for
  high rigidities}\label{sec:num}

\subsection{Numerical set up}
A Monte Carlo strategy is adopted to measure the diffusion
coefficients by integrating a large number of particle trajectories in
given turbulent magnetic field configurations.Averages are then
performed and statistical values of the diffusion coefficients
deduced.  The numerical set up is presented hereafter: we first
discuss the construction of the magnetic field, then the integration
of particle motion from Lorentz-Newton equation, and finally
the estimates of the diffusion coefficients.

The total magnetic field is expressed as
$\boldsymbol{B}=\boldsymbol{B_0}+\boldsymbol{\delta B}$ as before. The
regular field is oriented along $z$, and $\boldsymbol{\delta B}$ is
assumed to be isotropic in the three dimensions.  An algorithm similar
to Giacalone \& Jokipii (1999) is used to construct the turbulent
component of magnetic field $\boldsymbol{\delta B}$ by summing over
plane wave modes ($N_{mod}$) with turbulent wavelengths extending from
$L_{min}=1\equiv 2\pi/k_{\rm max}$ to $L_{max}\equiv 2\pi/k_{\rm
  min}$, the power spectrum following a truncated power law between
$k_{\rm min}$and $k_{\rm max}$. In detail
 
\begin{eqnarray}
  \boldsymbol{\delta B}(\boldsymbol{x}) = \sum_n G_n(k_n)
  \boldsymbol{\xi_n}\cos
  \left(\boldsymbol{k_n}\cdot\boldsymbol{x}+\beta_n\right).
\end{eqnarray}   
With Fourier modes of amplitude $G_n$, and wave vectors
$\boldsymbol{k_n}=\frac{2\pi}{L_n}\boldsymbol{e_k}$ isotropically
distributed, the unitary vector $\boldsymbol{\xi_n}$ is perpendicular
to $\boldsymbol{k_n}$ in order to ensure that
$\boldsymbol{\nabla}\cdot\boldsymbol{\delta B}$=0, and $\beta_n \in
[0,2\pi]$ represents the random phase. The power spectrum is
normalized by the turbulence parameter $\eta$ introduced earlier such that
$\langle\boldsymbol{\delta B}^2\rangle=B_0^2\eta/(1-\eta)$.  For
definiteness, the mode amplitudes are constructed according to a
Kolmogorov cascade with logarithmic spacing between wavenumbers: $G_n
\propto k_n^{-5/3}$. We note that the details of the inertial range of
the turbulence are not important because we are interested in the
scattering properties at large rigidities, when the particle Larmor radius
$\bar r_{\rm L}$ is larger than all turbulent length-scales.  For a
detailed presentation of the numerical turbulent magnetic field
construction, the reader is referred to Section 2.B of Casse et
al. (2002) and Section 3 of Giacalone \& Jokipii (1999).
   
Several tests of the dynamic range of turbulence $L_{\rm max}/L_{\rm min}$ and
the magnetic wave-modes $N_{\rm mod}$ were performed. The main
difficulty is that the scattering timescales increase
as a square of particle rigidity. For large rigidities, it is thus
difficult to preserve the accuracy with time when achieving particle
diffusion together with a realistic magnetic field model. To develop
 a simulation that operates over a few scattering times,
one needs to achieve an integration time of at least $100 \rho \bar
r_{\rm L}/c$, as in our simulations. One must also strike a
compromise with the number of plane wave modes to save computational
time; values of order $200-300$ have emerged as a satisfactory compromise
between accuracy and calculation time. To save computational
time, and because the small scales of the turbulent cascade are of
little influence, the dynamic range has been shortened to
$L_{\rm min}/L_{\rm max}=0.1$. Tests performed with a larger dynamic range
have provided similar results; the highest accuracy is obtained when
modes are concentrated on the largest scale. It is
explained physically by the high energy particles interacting only with
the largest magnetic structures. 
  
Particle motion is solved using the Lorentz-Newton equation of motion
that preserves its energy, hence its Lorentz factor $\gamma$
\begin{eqnarray}
  \frac{{\rm d}\boldsymbol{v}}{{\rm d}t }  =  
  \frac{q}{\gamma m c}\boldsymbol{v} \times \left(\boldsymbol{B_0}+ 
\boldsymbol{\delta B}\right)\ . \label{eq:Lorentz-Newton}
\end{eqnarray}
At this point, we define the numerical rigidity $\rho' \equiv 2 \pi
\bar r_{\rm L} / L_{\rm max}$, which differs from the previous
physical definition by a numerical factor of order unity, as discussed
earlier. The exact relation between $\ell_{\rm c}$ and $L_{\rm
  max}/2\pi$ depends on the dynamic range and the power-law index of
turbulence. In the following, the conversion factor between both
rigidities is derived using $\ell_{\rm c} \simeq 0.1 L_{\rm max}$, a
good approximation for a Kolmogorov-type spectrum.
    
The numerical integration of Eq.~(\ref{eq:Lorentz-Newton}) is
performed using a Bulirsch-Stoer schema (Press et al.1986). Once a
large number of particle trajectories were calculated and stored,
statistical averages instead being performed. Given the number of particles
$N_{\rm p}$ for each field realization and the number of field
realizations $N_{\rm field}$, the diffusion tensor coefficient $(i,j)$
is evaluated as
\begin{eqnarray}
  D_{ij}(t) & =
  & \frac{1}{2N_{\rm p}N_{\rm field}}\sum_{n=1}^{N_{\rm
      field}}\sum_{k=1}^{N_{\rm p}}
  \frac{(x_i(t)-x_i(t_0))(x_j(t)-x_j(t_0))_{n,k}}{t-t_0} \nonumber\\
  & = &  \frac{\left\langle \Delta x_i \Delta x_j\right\rangle}{2\Delta t}.
\end{eqnarray}
The average is performed over different particle trajectories and
different field realizations. For each value of $\rho$, we take
$N_{\rm field} \times N_{\rm p}=10^3$ different trajectories with
random initial velocity directions. The asymptotic value for $t \to
\infty$ (plateau) is roughly constant and defines the actual diffusion
regime. It gives the diffusion coefficient as $D_{ij}$ as $t \to
\infty$, precisely when $t \gg \nu_{\rm s}^{-1}$. This method of
coefficient estimation appears precise enough for an integration
involving $10^3$ particles. A complementary technique consists of
evaluating time correlations between velocities over particle
trajectories. With 1000 particles in the transport regime studied
here, this method is affected by numerical noise for the
velocity correlation function, hence is not presented.

Two different cases were investigated numerically: a pure
turbulence situation ($B_0=0$) and a strong turbulent case with
$\delta B \gg B_0$. Results are presented in the following
sub-sections.
  
\subsection{Pure turbulence $B_0=0$} \label{subsec_Diso}
  
These simulations were performed to test the correctness and
accuracy of the code. On theoretical grounds (see the appendix of
Casse et al. 2002, Aloisio et al. 2004, Pelletier et al. 2009) and
previous numerical works (Parizot 2004), we expect the diffusion
coefficient to evolve as the square of energy (e.g. rigidity) when
$\rho'\gg1$.
   
\begin{figure}
\begin{center}
\includegraphics[width=0.45\textwidth]{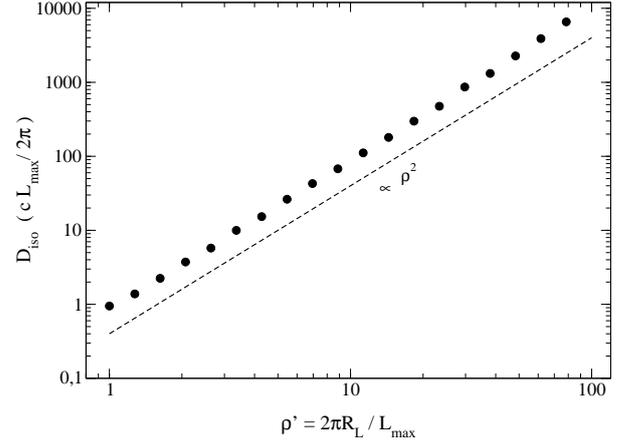}
  \caption{The diffusion coefficient variation is plotted in units of
    $c L_{\rm max} / (2 \pi)$ as a function of rigidity $\rho'$ in
    pure turbulence ($B_0=0$ or $\eta=1$). The dashed line is drawn as
    a reference for a scaling $D_{iso} \propto \rho'^2$.  For
    $\rho'>1$, $D_{iso}$ is indeed proportional to $\rho^2$.  }
\label{DISO}
\end{center}
\end{figure}

Here we set $\eta=1$ and $\boldsymbol{\delta B}$ isotropically
distributed by construction, so that the three space directions are
equivalent.  The equivalence of the three directions was
numerically verified in our simulations.  The diffusion coefficient is
evaluated as
\begin{equation}
  D_{\rm iso}  =  \frac{\left\langle \Delta x^2\right\rangle+
    \left\langle \Delta y^2 \right\rangle+ \left\langle \Delta z
      ^2\right\rangle}{6 \Delta t}\ . 
\end{equation}
Figure \ref{DISO} shows numerical values calculated for $\rho'$ going
from 1 to 100. The diffusion coefficient is plotted in units of $c
L_{\rm max} / (2 \pi)$ as a function of rigidity $\rho'$. A power law is
observed for $1<\rho'<100$, as predicted by the theory, namely
$D_{iso} \propto \rho'^2 \propto \epsilon^2$. One may be able to discern a
slight deviation at $\rho'$ close to 100. This purely numerical effect
disappears when taking a larger number of magnetic wave-modes on scales close to
$L_{\rm max}$ by defining $L_{\rm min}/L_{\rm max} \sim 1$. We retain however the
current field configuration, taking this effect into account when
interpreting the results.
  
\subsection{Weak mean field $B_0<\delta B$ }
  
We now consider the case where a constant weak mean field $B_0$ along
the $z$ direction is present.  In this case, two different diffusion
coefficients are defined $D_{\parallel}=D_{zz}$ and
$D_{\perp}=(D_{xx}+ D_{yy})/2$. Overall, we explored five different
levels of turbulence $\eta=\{0.5,0.9,0.99,0.999,0.9999\}$, spanning five
orders of magnitude in $\delta B^2/B_0^2$. The rigidity $\rho'$
ranges from 1 to 100 for each value of $\eta$. At each calculation
point $\{\eta,\rho\}$, the coefficients are evaluated by averaging over
$10^3$ particles (10 particles $\times$ 100 field realizations), as
before.
  
\begin{figure}
\begin{center}
\includegraphics[width=0.45\textwidth]{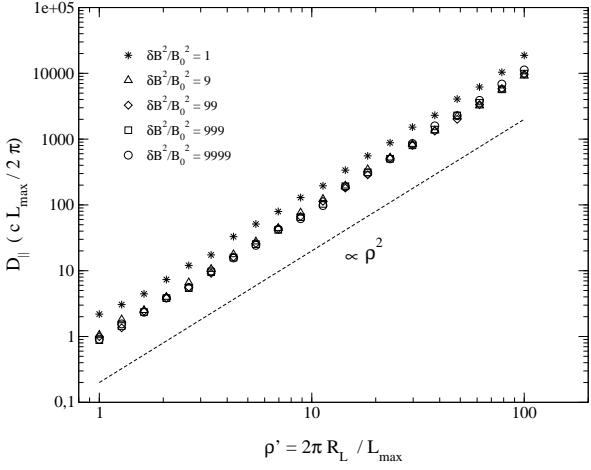}
  \caption{The parallel diffusion coefficient $D_{\parallel}$ plotted
    in units of ($c L_{\rm max} / (2 \pi)$) as a function of $\rho'$
    for different degrees of turbulence $\delta B^2 / B_0^2 \in
    [1,9999]$. Here $D_{\parallel}\propto \rho^2$, as in the case of purely
    isotropic turbulence without mean field. As long as $\delta B^2 /
    B_0^2\gg1$, the strength of the turbulence does not influence the
    normalization of $D_{\parallel}$.}
\label{dpar}
\end{center}
\end{figure}

As shown in Fig.~\ref{dpar}, the parallel diffusion coefficient
retains the same dependence on rigidity as in pure turbulence,
$D_{\parallel} \propto \rho'^2$. For $\eta>0.5$, the turbulence level
has almost no influence on the value of $D_{\parallel} \simeq 0.9
c\bar r_{\rm L}\rho'$. Therefore as expected, the mean field, as long
as it remains weak enough, has no influence on the diffusion of
particles along its direction.

\begin{figure}[h]
\begin{center}
\includegraphics[width=0.45\textwidth]{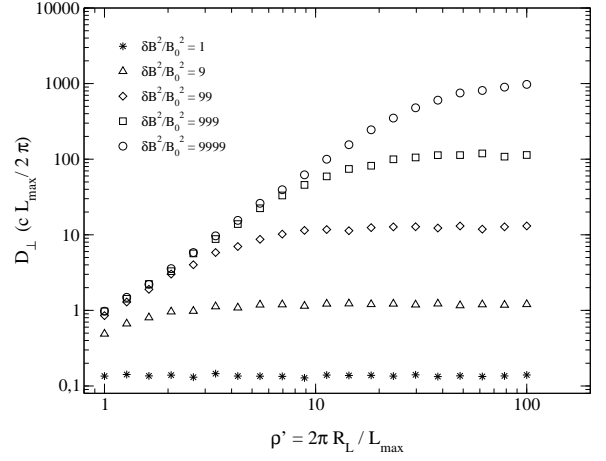}
  \caption{The transverse diffusion coefficient $D_{\perp}$ plotted in
    units of $c L_{\rm max} / (2 \pi)$ as a function of $\rho'$ for
    different degrees of turbulence $\delta B^2 / B_0^2 \in
    [1,9999]$. The diffusion coefficient saturates at $\rho'\sim
    \delta B / B_0$. Below this value, its behavior is similar to the
    parallel diffusion coefficient. Beyond $\rho'$, its value becomes
    independent of particle rigidity. }
\label{dperp}
\end{center}
\end{figure}

The picture is different for the transverse coefficient
when the particle rigidity becomes large. In Figure \ref{dperp}, the
simulated transverse diffusion coefficient is plotted as a function of
rigidity $\rho'$ for different degrees of turbulence. In each case, its
value saturates to a constant value when $\rho' \sim \delta B /
B_0$. This value behaves proportionally to the turbulence degree; in
detail, $D_{\perp} \simeq 0.13 c (L_{\rm max}/2\pi)\delta B^2/B_0^2$,
in excellent agreement with our theoretical prediction from
Eq.(\ref{eq:dperp_th}). Individual particle trajectories reveal a
weakly perturbed helical path when $\rho'>>1$. Therefore, a strong
small-scale turbulence acts as a collection of small-scale scattering
centers, each producing a small deflection.

According to the theory, $D_{\parallel}$ is the limit of a function
$c^2g_{\parallel}(t)/3$ as $t \rightarrow \infty$, precisely
as $t>t_s$, the function being
\begin{equation}
  g_{\parallel}(t) = \frac{1-e^{-\nu_{\rm s}t}}{\nu_{\rm s}} \ .
\end{equation}
In a similar way $D_{\perp}$ is the limit of a function $c^2g_{\perp}(t)/3$
as $t \rightarrow \infty$, in addition to when $t>t_{\rm s}$, the function being
\begin{equation}
g_{\perp}(t) = \frac{\nu_{\rm s}}
{\Omega_{0}^2 + \nu_{\rm s}^2}\left\{1-e^{-\nu_{\rm s}t}\left[\cos
\left(\Omega_{0} t\right) -\frac{\Omega_{0}}{\nu_{\rm s}} 
\sin\left(\Omega_{0}t\right)\right]\right\} \ .
\end{equation}

\begin{figure}[b]
\begin{center}
\includegraphics[width=0.45\textwidth]{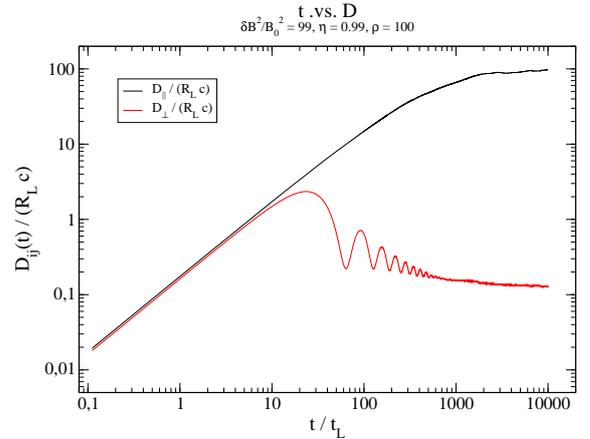}
  \caption{Transition toward parallel and perpendicular
    diffusion. Before reaching its asymptotic value for $t > \tau_{\rm
      s}$, the transverse diffusion rate decreases as in a
    sub-diffusive regime.}
\label{EDCs}
\end{center}
\end{figure}
 
The numerical simulation reproduces these types of behavior, although the transverse
evolution departs slightly from the above formula before
reaching the scattering time $\tau_{\rm s}$.  Nevertheless, the
agreement between the theory and the numerical simulation holds during
the linear growth at the beginning of the evolution and when the
evolution approaches the asymptotic behavior.  The numerical results
confirm the theory we proposed in the previous section for the
asymptotic regime. The scattering time is clearly the time beyond
which spatial diffusion takes place. We can also note that there is a
sub-diffusion regime before the settlement of the transverse diffusion
regime.

\begin{figure}
\begin{center}
\includegraphics[width=0.45\textwidth]{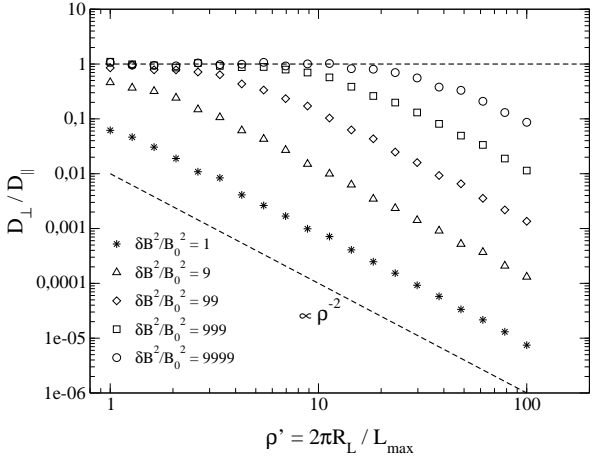}
  \caption{Anisotropy ratio $D_{\perp}/D_{\parallel}$ as function of
    $\rho'$ for different levels of turbulence $\delta B^2 / B_0^2 \in
    [1,9999]$, as indicated by the various symbols. The dashed line
    provides a guide for a ${\rho'}^{-2}$ scaling.}
\label{RDC}
\end{center}
\end{figure}

The anisotropy ratio $D_{\perp} / D_{\parallel}$ can be seen in Figure
\ref{RDC} as a function of $\rho'$. When the turbulence level $\eta$ is
close to 1 and $\rho'$ is not too large, the transport appears
isotropic $D_{\perp}/D_{\parallel} \simeq 1$.  At higher rigidities,
its behavior follows the law $\propto \rho'^{-2}$ for all turbulence
levels, illustrating the saturation of the transverse coefficient and in
agreement with the theoretical prediction.

\subsection{Comparisons with previous results}
Transverse diffusion at high rigidity, as
far as we know, has been poorly studied in the
literature. However, we can compare our results with several previous
numerical and theoretical studies with different limits.

The seminal study of Giacalone\& Jokipii (1999) focused on the
propagation of mildly relativistic particles ($E=1$MeV to 1 GeV) in
the interplanetary magnetic field ($\delta B^2 \sim B_0^2$). Their
simulations provided results for $\rho \leq 1$ and $\eta \leq
0.5$. However, they also performed several simulations in which
the particle energy and the coherence length remained fixed, while the
turbulence level was varied. In particular, they examined the case
$r_{\rm L,0}/ \ell_{\rm c} = 10$ for moderate values of $\delta B^2/
B_0^2$ (Fig.~6 of their paper) in which $D_{\perp}/D_{\parallel}$
is plotted as a function of $\lambda_{\parallel}/r_{\rm L,0}$
($\lambda_\parallel$ denoting the mean free path in the parallel
direction). By inspecting their figure, one can see that they varied
$\delta B^2/ B_0^2$ from 0.05 to 30. As a result, they found a
classical scattering theory scaling but no physical explanation was
proposed. Strictly speaking, the classical theory is valid only for
weak turbulence ($\delta B^2\ll B_0^2$), which clearly does not apply
to those simulations. The present theoretical framework provides a
clear explanation of this result, which we confirmed with
additional detailed numerical simulations. It is found, for instance, 
that particles with large
rigidities do not interact directly with the magnetic field lines but
experience an overall magnetic topology dominated by the mean field
with ``infinite'' coherence length.  As a result, the particles execute
regular orbits around $B_0$ and undergo random deflections on the
coherence length-scale.
  
The simulations of Casse et al. (2002) investigated weak as well
as strong turbulence regimes where $\delta B^2 / B_0^2 \in [0.1, 99]$. An
FFT algorithm was used to construct the magnetic field. For $\rho'>1$,
these authors found evidence of anisotropic scattering
$D_{\perp}/D_{\parallel}<1$ for all turbulence levels. However, only three
simulations points were computed in the high rigidity range and the
estimate of the power law slope was inaccurate. Nevertheless, a
reasonable agreement is obtained when comparing values of
$D_{\parallel}$ and $D_{\perp}$ with the present results.
  
Parizot (2004) presented simulations of particle propagation in
pure isotropic turbulence.  The results in the regime $r_{\rm
  L}\gg\ell_{\rm c}$ leads to a diffusion coefficient with a quadratic
scaling, $D \propto E^2$, in agreement with our results from Sec~\ref{subsec_Diso}.
 \begin{figure}
\begin{center}
\includegraphics[width=0.45\textwidth]{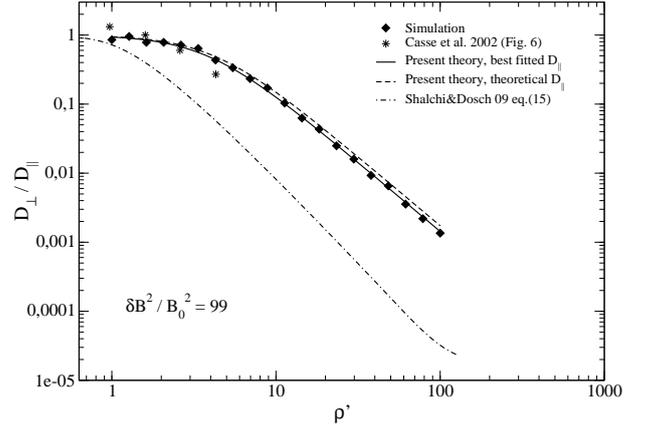}
  \caption{Ratio $D_{\perp}/D_{\parallel}$ as a function of $\rho'$ for
    $\eta=0.99$, compared to theorectical predictions and other
    numerical simulations. Filled diamonds: our simulation
    results. Star symbols: results from Casse et al. 2002. Solid
    curve: present theoretical prediction with best-fit
    $D_{\parallel}$ from simulations (see Fig.~\ref{dpar}). Dashed
    curve: present theoretical prediction with analytical
    $D_{\parallel}=c^2/(3 \nu_{\rm s})$. Dot-dashed curve: analytical
    prediction from Shalchi \& Dosch (2009), their Eq.~(15).}
\label{CRD}
\end{center}
\end{figure}
  
Shalchi \& Dosch (2009) derived an analytical expression for the
diffusion anisotropy ratio $D_{\perp}/D_{\parallel}$ in the framework
of a non-linear guiding centretheory. They assume an isotropic
turbulence $\boldsymbol{\delta B}$ with a mean field
$\boldsymbol{B_0}$. No assumption was made about either the level of
turbulence or about particle energy, hence their result should be
valid for any particle rigidity and turbulent field strength. An
expression of $D_{\perp}/D_{\parallel}$ [Eq.~(15) in their work] that
depends on two parameters was obtained.  The first parameter
corresponds to the ratio of the mean free path ($\lambda_{\parallel}$)
along the mean field direction to the coherence length $\ell_{\rm c}$
of the turbulent field. The second parameter is the turbulence level
$\delta B^2/ B_0^2 = \eta/(1-\eta)$. Shalchi \& Dosch (2009) thus find
that the transport becomes highly anisotropic, meaning
$D_{\perp}/D_{\parallel}\ll1$ when $\lambda_{\parallel}/\ell_{\rm
  c}\gg1$ and/or $\delta B^2/ B_0^2$ is not too large (see Figs.~1 and
2 of their work). Therefore, our present conclusions agree with theirs,
at least at a qualitative level. A detailed comparison would require us to
define $\lambda_{\parallel}$ as a function of $\rho'$, which could be done by
using our results of $D_{\parallel}$ for which $\rho'=(4 \pi
\eta /30)^{1/2} (\lambda_{\parallel} / \ell_{\rm c})^{1/2}$. With this
substitution, we can directly compare their predictions to our
results. In Fig.~\ref{CRD}, we plot the ratio of diffusion
coefficients as a function of $\rho'$ from our numerical simulations
and compare these results to both the predictions of Shalchi \& Dosch
(2009) and the theoretical model developed in Sec.~\ref{sec:theor}.
Good agreement is found between the simulation results (diamond
symbols) and our theory (solid curve and dashed curve); however, the
predictions of Shalchi \& Dosch (2009) disagree with the numerical
results, increasingly so as the rigidity increases. In particular,
their analysis predicts a scaling with a slope $-2.4$ instead of 
the value of $-2$ observed here. 
Repeating the same comparisons for each simulated
value of $\delta B^2 / B^2$, we were unable to find agreement between
the predictions of Shalchi \& Dosch (2009) and our simulations; the
predicted values always lie below the numerical results, with a
different power-law scaling, comprised between -2.5 for $\delta B^2/
B_0^2 = 1$ and -2.4 for $\delta B^2 / B_0^2 =10^4$. At this point, it
could be argued that our definition of $\lambda_{\parallel}$ as a
function of $\rho'$ is inaccurate. However, on physical grounds, the
scaling $\lambda_{\parallel} \propto \epsilon^2$ when $\bar r_{\rm L}
\gg \ell_{\rm c}$ remains robust. Therefore, the discrepancy between
the power law scalings should not be affected by uncertainties in the
numerical prefactors. We think that the ``guiding center''  assumption 
in their work is questionable.

\section{Summary and some astrophysical applications}\label{sec:appl}

\subsection{Summary}

Our investigation of the diffusion process in small-scale 
turbulence with a mean field revealed that, despite its smallness, the mean 
field plays a role in transverse diffusion because the scattering frequency 
decreases like $\epsilon^{-2}$, whereas the Larmor frequency decreases like 
$\epsilon^{-1}$. Instead of finding a single diffusion coefficient 
that increases like $\epsilon^{2}$, we found an anisotropic diffusion with a 
transverse coefficient that reaches a limit value at large rigidities.
The theory we proposed is based on a single assumption, namely that the correlation 
time is much smaller than the scattering time, which is valid for both small and large 
rigidities. The only regime where the theory fails is for a rigidity close to 1 and 
a high turbulence level; however, the interpolation is obvious. 
The theory allows us to derive a correct pitch-angle scattering 
rate and a correct parallel diffusion coefficient for every rigidity. It
provides a transverse diffusion coefficient similar to the classical scattering theory 
formula, despite the arbitrary level of turbulence, which is a correct result for large rigidity.
At low rigidity, the present theory is incorrect because it does not take into account the 
effect of field line wandering described in Casse et al. 2002.

\subsection{Particle transport in relativistic shock environments}
One major application of the diffusion theory in small-scale
turbulence is the transport of supra-thermal particles in the vicinity
of a relativistic shock. By crossing the shock transition,
electrons and protons reach more or less the same characteristic
energy $\langle\epsilon\rangle \sim \gamma_{\rm sh} m_p c^2$ as
revealed clearly by particle-in-cell simulations (e.g., Sironi \&
Spitkovsky 2011). There is a single plasma frequency $\omega_{\rm p*}
\sim \omega_{\rm pi}$, where $\omega_{pi}$ is the ion plasma frequency
in the upstream or unshocked plasma. This length-scale characterizes
the typical length scale of the microturbulence excited in the shock
precursor, as transmitted downstream of the shock transition and
viewed in the downstream rest frame. The generation
of short scale intense micro-turbulence is possible only at low
magnetizations of the upstream plasma (Sironi \& Spitkovsky 2011), where the
magnetization parameter $\sigma$ is here defined as the flux of
magnetic energy crossing the shock over the flux of matter energy,
$\sigma \equiv B_0^2 \sin^2 \theta_B/4\pi \rho_{\rm u} c^2$ (where
$\theta_B$ is the angle of the background magnetic field with the shock
normal, and $\rho_{\rm u}$ the unshocked plasma mass density). However,
this same level of magnetization also permits the
efficient acceleration of particles through a first-order Fermi
process at the shock front (Lemoine \& Pelletier 2010, 2011). 
For larger magnetizations -- the exact level depending on the shock
Lorentz factor, see the above references -- the Fermi process cannot
develop because of a lack of efficient scattering in the microturbulence
(Lemoine et al. 2006, Niemiec et al. 2006, Pelletier et al. 2009). In
brief, the development of the Fermi process hinges on the development
of micro-turbulence, which itself requires (in the absence of external
sources of turbulence) a sufficiently low magnetization level. The
situation in which particles are accelerated is by far the most
interesting as it should produce directly observable signatures, in
the form of radiation and possibly neutrinos.

The transport properties of these accelerated particles is then
directly governed by the parallel and perpendicular diffusion
coefficients in the limit of large rigidity, as discussed
above. We assume that the microturbulence has a typical length-scale
 close to $\delta_*=c/\omega_{\rm p*}$ and that a fraction
$\epsilon_B$ of shock dissipated energy is converted into
electromagnetic turbulence, i.e.
\begin{equation}
\frac{\langle\delta B^2\rangle}{8\pi} = 2\epsilon_B\gamma_{\rm sh}^2
\rho_{\rm u} c^2 \ ,
\end{equation}
where the rigidity of shock accelerated particles of energy $\epsilon$ is given by
\begin{equation}
  \rho \approx \epsilon_B^{-1/2} \frac{\delta_*}{\ell_{\rm c}}
\frac{\epsilon}{\langle\epsilon\rangle}\ .
\end{equation}
Current simulations indicate values of $\epsilon_B\sim 0.01-0.1$, hence
$\rho>1$ and all the more so at high energy.

In this regime, the perpendicular diffusion coefficient that we
discussed in the previous section becomes particularly relevant, as
the mean magnetic field is mostly perpendicular to the shock normal in
the downstream frame, since the transverse components (relatively to
the shock normal) are increased by $2\sqrt{2}\gamma_{\rm sh}$, while
the parallel component remains the same as in the upstream
frame. Therefore, perpendicular diffusion at high rigidity plays an
essential role in the transport of particles in the downstream flow of
a relativistic shock.

We consider the diffusive behavior of particles in the downstream rest frame. In this frame the 
shock front appears to move away with velocity $V_{\rm shock} \simeq c/3$. Achieving Fermi cycles requires the particle  to return to the shock front. The return time is then  measured by identifying shock front with the  particle mean displacements
\begin{equation}
  {c \over 3}t_{\rm ret} = \sqrt{2 D_{\perp} t_{\rm ret}} .
\end{equation}
Therefore $t_{\rm ret}=18D_{\perp}/c^2$ and Fermi cycles are possible until $t_{\rm ret}$ is neither large nor too short. While the first case is constrained by confinement in the acceleration site, the second one is related to the  diffusive approximation that is valid only when $t_{\rm ret} \geq \tau_s$.  Using the second limit to constrain diffusive returns, one obtain $D_{\perp}/ D_{\parallel} \geq 1/6$,  equivalent to $\nu_s \geq \sqrt{5} \Omega_0$ when $D_{\perp}$ is replaced by its expression from Eq.~(\ref{eq: coeff-dperp}). Fiducial values for a relativistic shock in the interstellar medium provide an energy limit $E_{\rm lim} \sim 10^{19}$eV. This limit is somewhat irrelevant because $t_{\rm ret}\gg R_{\rm acc}/c$ at this energy, where $R_{\rm acc}/c$ is the shock dynamics timescale. Hence, the returns appear to be efficient when the condition $\tau_s < t_{\rm ret} \ll R_{\rm acc}/c$ is satisfied. Further investigation would require us to solve a kinetic equation taking into account acceleration, scattering and energy losses processes. Diffusion coefficients obtained in this work may be relevant to providing more realistic results. Previous works assumed Bohm diffusion or isotropic pitch-angle scattering.

Detailed discussion of the performance of the
relativistic Fermi process is beyond of the scope of the
present paper and is left to future work. 

In certain astrophysical settings, the transverse diffusion may play a
key role in the transport of particles upstream of a relativistic
shock, most particularly if the shock propagates in a wind with a
dominant toroidal field at large distances.  These circumstances can be
encountered in particular when a gamma-ray burst explodes in the wind
of the progenitor, or at the termination shock of a pulsar wind. 

\subsection{High-energy cosmic rays}
The above result about transverse diffusion has a broader application
than Fermi acceleration at shocks, as it governs the confinement
properties of any relativistic flow containing a small-scale
turbulence, where ``small'' is measured relatively to the Larmor
radius of the test particles propagating in this flow.  This concerns
in particular the propagation of very high-energy cosmic rays in our
Galaxy. Assuming a coherence length of interstellar turbulence $\ell_{\rm c}
\sim 10-100\, pc$, a mean field intensity of $3 \, \mu G$
approximately and a turbulent field of the same order, the rigidity of
particles of energy $E$ is given by $\rho \simeq 2 (E/10^{17}\,{\rm
  eV})(\ell_{\rm c}/10\,{\rm pc})^{-1}(\bar B/5\,\mu{\rm G})^{-1}$,
while the Larmor radius $r_{\rm L}\simeq 20\,{\rm
  pc}\,(E/10^{17}\,{\rm eV})(\bar B/5\,\mu{\rm G})^{-1}$.
  Assuming $\eta \simeq 0.5$ in the Galaxy and using 
   Eq.~(\ref{eq: coeff-dperp}), the perpendicular mean free path
  is then of order $\lambda_{\perp} \sim 6 \,{\rm pc}$ 
  with these values of energy and magnetic field. This implies
that the escape, or transport across the disk magnetic field of
particles of energy $\geq 10^{17}\,$eV is governed by the
perpendicular diffusion in the high rigidity regime discussed
above. Quite interestingly, this energy range presumably corresponds
to the transition between the Galactic and extragalactic cosmic-ray
components in the all-particle spectrum. 

Finally, one could mention another application of the present
discussion, to the field of magnetic reconnection. There, transverse
diffusion likely plays a role in the control of particle diffusion
across the field lines with small-scale turbulence being associated with the
dissipation of magnetic energy. The reconnection rate depends on two
fundamental parameters (Lyutikov \& Uzdensky 2003): magnetization and
the Lundquist number that involves diffusion across field lines. In
general, one assumes Bohm diffusion for simplicity but the present
work provides the grounds for a more accurate estimate.

\appendix

\section{Average of the time ordered exponential}

We solve the differential equation $\dot {\mathbf u} =
\boldsymbol{\hat {\tilde \Omega}}\cdot \mathbf{u}$ in successive
iterations that leads to a Dyson series, the average of which is
composed of products of the form
\begin{equation}
 \boldsymbol{ \hat A_{2p}(t)} = \int_0^{t} {\rm d}t_1\int_0^{t_1} {\rm d}t_2\ldots\int_0^{t_{2p-1}}
  {\rm d}t_{2p}\,
  \langle\boldsymbol{\hat {\tilde\Omega}(t_1)} \cdot 
\boldsymbol{\hat {\tilde \Omega}(t_2)}\ldots 
\boldsymbol{\hat {\tilde \Omega}(t_{2p})}\rangle \ ,
\end{equation}
which can be compared to Eq.~(\ref{eq:Dyson-series}).

For a Gaussian process, each average of order $2p$ products can be
divided into a sum of $p$ products of second-order moments, the sum
containing $(2p-1)!!$ terms. We assume a stationary random process such
that the second order moment is an even function of the time
difference. In the white noise limit, the ``nested" and ``crossed"
averages vanish, only the ``unconnected" averages remaining in the
expansion. Nested terms contain products of the form $\langle
X(t_i)X(t_l)\rangle\langle X(t_j)X(t_k)\rangle$ with $t_i\geq t_j\geq
t_k\geq t_l$, while crossed terms are of the form $\langle
X(t_i)X(t_k)\rangle\langle X(t_j)X(t_l)\rangle$ with $t_i\geq t_j\geq
t_k \geq t_l$. These terms vanish as the various delta functions
associated with the second order moments cancel each other as a result
of the time ordering in the upper bounds of the integrals. Thus, only
the unconnected average remains at each order
\begin{eqnarray}
  \boldsymbol{\hat A_{2p}(t)} &=& \int_0^{t} {\rm d}t_1\int_0^{t_1} {\rm
    d}t_2...\int_0^{t_{2p-1}} {\rm d}t_{2p}\,\nonumber\\
&& \quad \langle\boldsymbol{\hat {\tilde\Omega}(t_1)} \cdot \boldsymbol{\hat
    {\tilde \Omega}(t_2)}\rangle\ldots
  \langle\boldsymbol{\hat {\tilde \Omega}(t_{2p-1})}\cdot\boldsymbol{\hat {\tilde \Omega}(t_{2p})}\rangle \ .
\end{eqnarray}

We introduce the short-hand notation $\langle\boldsymbol{\hat
  {\tilde\Omega}(t_1)} \cdot \boldsymbol{\hat {\tilde
    \Omega}(t_2)}\rangle \equiv \boldsymbol{\hat C(t_1-t_2)} = 2\tau_{\rm c}
\delta(t_1-t_2) \boldsymbol{\hat C_0}$.  Then one can calculate $\boldsymbol{\hat A_{2p}(t)}$
 by recursion, starting from the last double integral in the
product
\begin{equation}
\boldsymbol{\hat A_{2p}(t)} =  \frac{(\tau_{\rm c} t)^p}{p!} \boldsymbol{\hat C_0^p} \ .
\end{equation}
We consider now the integral of the second order moment
\begin{equation}
\boldsymbol{\hat K(t)} \equiv \int_0^t {\rm d}t_1 \int_0^t {\rm d}t_2 
\,\langle\boldsymbol{\hat{\tilde\Omega}(t_1)} \cdot 
\boldsymbol{\hat {\tilde \Omega}(t_2)}\rangle = 2\tau_{\rm c} t \boldsymbol{\hat C_0} \ .
\end{equation}
Therefore
\begin{equation}
\boldsymbol{\hat A_{2p}(t)} = \frac{1}{2^p p!} \boldsymbol{\hat K^p(t)} \ ,
\end{equation}
hence summing all the terms of the series, 
\begin{equation}
\sum_{p=0}^{p=+\infty} \boldsymbol{\hat A_{2p}(t)} = \exp\left[\frac{1}{2} \boldsymbol{\hat K(t)}\right] \ .
\end{equation}  
Further details can be found in Frisch (1966) and Pelletier (1977).

\end{document}